\documentclass[journal]{IEEEtran}
\usepackage[T1]{fontenc}
\usepackage{lmodern}
\usepackage{stackengine}

\usepackage{amsmath}

\usepackage{stackengine}

\usepackage{wrapfig}
\usepackage{caption}
\usepackage[framemethod=tikz]{mdframed}
\usetikzlibrary{calc}
\usepackage{kantlipsum}
\begin{document}
%
\title{A Nano-Architecture for Fertility Monitoring via Intra-body Communication}
\fontfamily{lmss}\selectfont
%
%
%

\author{Shama Siddiqui, Anwar Ahmed Khan,  Qammer H. Abbasi, ~\IEEEmembership{Senior Member,~IEEE}, \\Muhammad Ali Imran,~\IEEEmembership{Senior Member,~IEEE} and Indrakshi Dey,~\IEEEmembership{Senior Member,~IEEE}

\thanks{S.~Siddiqui is with Department of Computer Science, DHA Suffa University, Karachi, Pakistan (Email: shama.siddiqui@dsu.edu.pk)}
\thanks{A.A.~Khan is with Department of Computer Science, Sindh Institute of Management \& Technology, Pakistan (Email: hod.cs.it@simt.edu.pk)} %
\thanks{I. Dey is with CONNECT, National University of Ireland, Maynooth (E-mail: indrakshi.dey@mu.ie)}
\thanks{Q. H. Abbasi \& M.A. Imran are with James Watt School of Engineering, Unviersity of Glasgow (E-mails: qammer.abbasi; muhammad.imran@glasgow.ac.uk)}
}


\maketitle

\begin{abstract}

Fertility monitoring in humans for either natural conception or artificial insemination and fertilization has been a critical challenge both for the treating physician and the treated patients. Eggs in human female can be fertilized when they reach the Fallopian tube from the upper parts of the reproductive system. However, no technology, till date, on its own could detect the presence of eggs in the Fallopian tube and communicate their presence to the consulting physician or nurse and the patient, so that the next step can be initiated in a timely fashion. In this paper, we propose a conceptual architecture from a communications engineering point of view. The architecture combines intra-body nano-sensor network for detecting Fallopian tube activity, with body-area network for receiving information from the intra-body network and communicating the same over-the-air to the involved personnel (physician/nurse/patient). Preliminary simulations have been conducted using particle based stochastic simulator to investigate the relationship between achievable information rates, signal to noise ratio (SNR) and distance. It has been found that data rate as high as 300 Mbps is achievable at SNR 45. Hence, the proposed architecture ensures transfer of information with near-zero latency and minimum energy along with high throughput, so that necessary action can be taken within the short time-window of the Fallopian tube activity.  

\end{abstract}


%
\IEEEpeerreviewmaketitle

\section{Introduction}
%
%
%
%
\IEEEPARstart{T}HE emerging developments in the domains of nano-sensors and intra-body communication promises to revolutionize the future healthcare through facilitating pre-diagnosis of impending medical conditions. To facilitate the innovative diagnosis and treatment options, nano-structured materials have been used to design miniature implantable bio-sensors. These sensors have capability of real-time data acquisition from various parts of the body \cite{ram}. Nano-structured materials appear ideal for intra-body sensing and communications in healthcare applications, due to their unique electronic, chemical, structural properties and large surface to volume ratio and their use has been recommended as disease detector, drug carrier, filler, filter and reaction catalyst. Some of the major evolving application areas of intra-body nano-sensing and communications include drug delivery, detection of toxic substances, micro-bacteria, viruses or allergens, early detection of cancer cells, etc.\cite{khan}. 

Due to the underlying challenges of using conventional Electromagnetic communications within the intra-body environment, Body-Centric Nano Networks (BCNNs)/Internet of Bio Nano things (IoBNT) have taken inspirations from the biological communication processes \cite{yang}. In this regard, the fields of Molecular communication (MC) has evolved, which deals with exchange of molecules by the transmitter and receiver. Also, the Tera-hertz (THz) band (0.1 - 10 THz) has been regarded as one of the best candidates for intra-body communication as it offers high reliability and negligible latency \cite{ghafoor}. The use of hybrid scheme that combines molecular and THz communication for intra-body applications has recently been proposed \cite{yang} for the timely information transmission about the ongoing physiological processes. 

In MC, transmitters, receivers and/or actuators are deployed in the human body using a predefined application specific topology and molecules are used as information carriers. The transmitters may encode the message as number, type or release time/pattern of the molecules; this message is subsequently sensed and decoded by the receiver nodes \cite{akyildiz}. The THz communication for intra-body applications are encouraged due to offering high bandwidth and low latency. Due to the higher speed and low latency of THz communication as compared to the MC, it is often preferred to be used over long distances within the intra-body. Also, the information from the intra-body network is transmitted to the on-body nano-devices using THz link, which is subsequently forwarded to a cloud-computing platform via an IoT Gateway, which further connects to the experts or emergency service providers for the appropriate action to be taken. 

As with other biological processes which occur at microscopic scale inside the human body, it has been a serious challenge to monitor the ovulation in order to ensure successful conception using natural or artificial methods \cite{ali}. Ovulation refers to the process when egg/s release from the female ovaries into the Fallopian tubes and begin travelling to the uterus. In general, the eggs remain in Fallopian tubes for up to 24 hours where they can be fused with sperm. At this time, for the natural conception to occur, it is crucial that the sexual intercourse is timed near ovulation so that the male sperm could swim up to the Fallopian tube and attach itself with one of the eggs. The duo then attaches to the endometrium (uterine lining) and proceeds to the later stages of pregnancy. On the other hand, in case the egg could not fertilize when in the Fallopian tube, the female body sheds its along with the endometrium via menstrual blood.  

At present, the females trying to conceive, have to rely both on estimates and ultrasound scans to monitor their fertility. Ovulation approximately takes place after 13-15 days before the start of each menstrual period. When in ovaries, the eggs remain in sacs referred as follicles. The presence and size of follicles in ovaries is detected using conventional pelvic and/or trans-vaginal ultrasound scans \cite{luo}. As a follicle reaches approximate size of 18-28 mm, it is considered ready for ovulation. When the eggs are released into the Fallopian tubes, the available non-invasive tracking methods (such as measuring Luteininzing Hormone, and maintaining Basal Body Temperature Chart) provide no guarantee to detect the presence of egg in real time \cite{marcinkowska}. Furthermore, the recent introduction of fertility monitoring mobile applications which rely on parameters such as dates of menstrual cycle, basal body temperature and presence of cervical mucus also cannot track the accurate time when eggs are present within the Fallopian Tube as they just function on tracking the symptoms which may be misleading \cite{hamper}. As a result, the probability of missing the ovulation increases even after going through a series of ultrasounds and diagnostic tests (urine and blood for LH monitoring), which in turn  increases probability of failed attempts of natural conception or 
Intrauterine insemination (IUI) and In-vitro fertilization (IVF) treatments. Therefore, in this work, we propose to use intra-body hybrid communication architecture integrated with nano-sensors/nano-devices for detecting presence of eggs in Fallopian tube in the real-time. Here, hybrid communication refers to the combined use of MC and THz communication to realize the advantages of both paradigms. 

\begin{figure*}[!htb]
\centering
\includegraphics[scale=0.75]{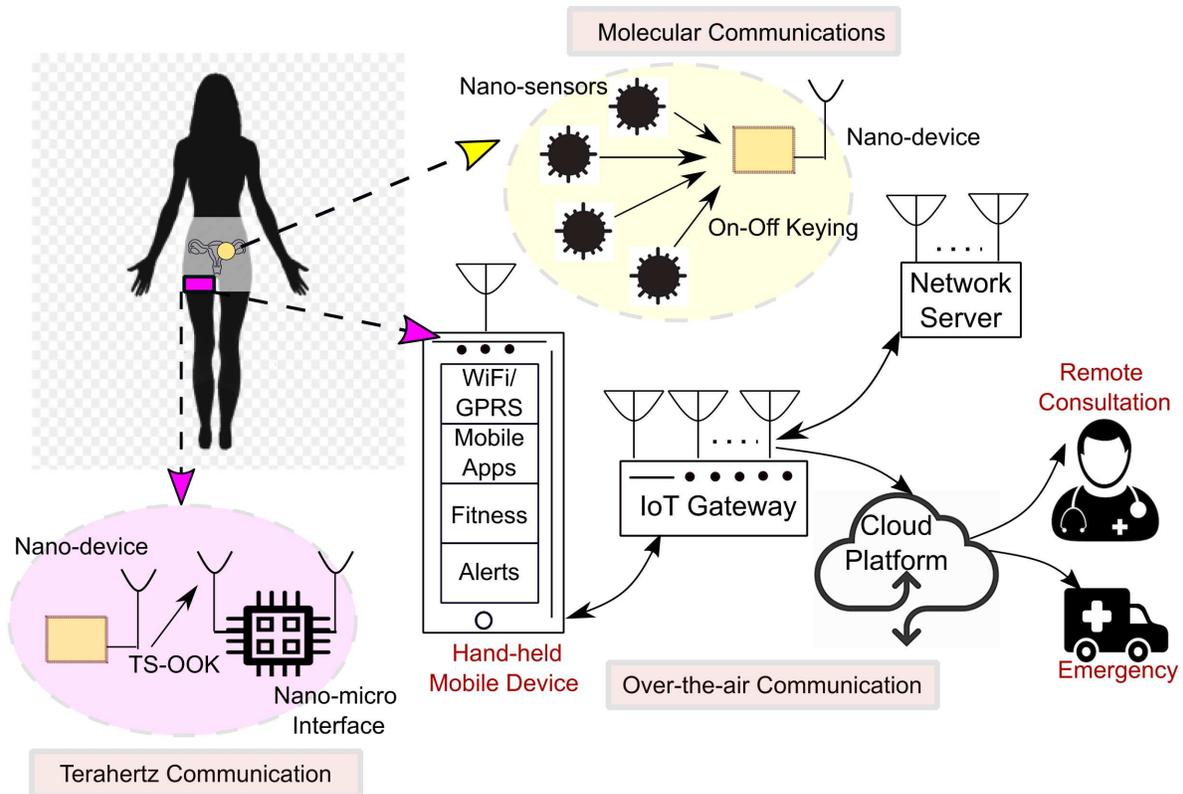}
\vspace{-115mm}
\caption{Overall architecture integrating intra-body, body-area and extra-body networks.}
\label{fig:Architecture}
\end{figure*}

To the best of our knowledge, no previous study has proposed to use intra-body communication techniques for monitoring the human fertility. We believe that acquiring real-time data from Fallopian tubes will provide precise information about the presence of eggs. Hence, the novel contribution of the present article is to develop an end-to-end architecture that integrates intra-body molecular and THz communication with body area network (BAN) and extra-body IoT backhaul network. It is proposed that the nano-sensors are implanted within the Fallopian tubes and transmit alert signals to the patient and physicians as soon as an egg is released. It is expected that the proposed architecture shall aid the patients and medical practitioners to monitor the fertility and precisely time the sexual intercourse for the possible natural conception via fertilization of eggs. 

 \section{Proposed Architecture}
We propose an end-to-end communication architecture to integrate the intra-body molecular and THz communication schemes with the BAN and extra-body IoT backhaul communication networks, for the purpose of efficiently, timely and reliable fertility monitoring. Both molecular and THz communications are safe for the transmission within human body and no hazardous effect has been shown in the previous studies \cite{saeed}. Since the system is based on collecting information directly from the Fallopian tubes, the delay of presently existing ultrasound and diagnostic testing solutions shall significantly reduce. Assisted diffusion based MC shall be used for collecting information from nano-sensors deployed within the Fallopian tubes. In addition, we also propose to use THz communication for the transmission of information between intra-body and on-body devices. Finally, the on-body device will forward information over-the-air to hand-held device, which could in turn send alert to the remote locations. 

The entire communication architecture illustrating transmission from intra-body nano-sensor nodes to the on-body device, hand-held device and remote server has been illustrated in Fig. \ref{fig:Architecture}. A molecular nano-network is deployed within the Fallopian tube of a monitored female individual. It consists of several nano-sensors that act as biological transmitter deployed in different section of the Fallopian tube. These sensors detect the presence or absence of newly released eggs in the tube. The sensor observations are then modulated and transmitted to an implantable graphene-based nano-device placed strategically at the end of the tube. 

\begin{figure*}[!htb]
\centering
\includegraphics[scale=0.8]{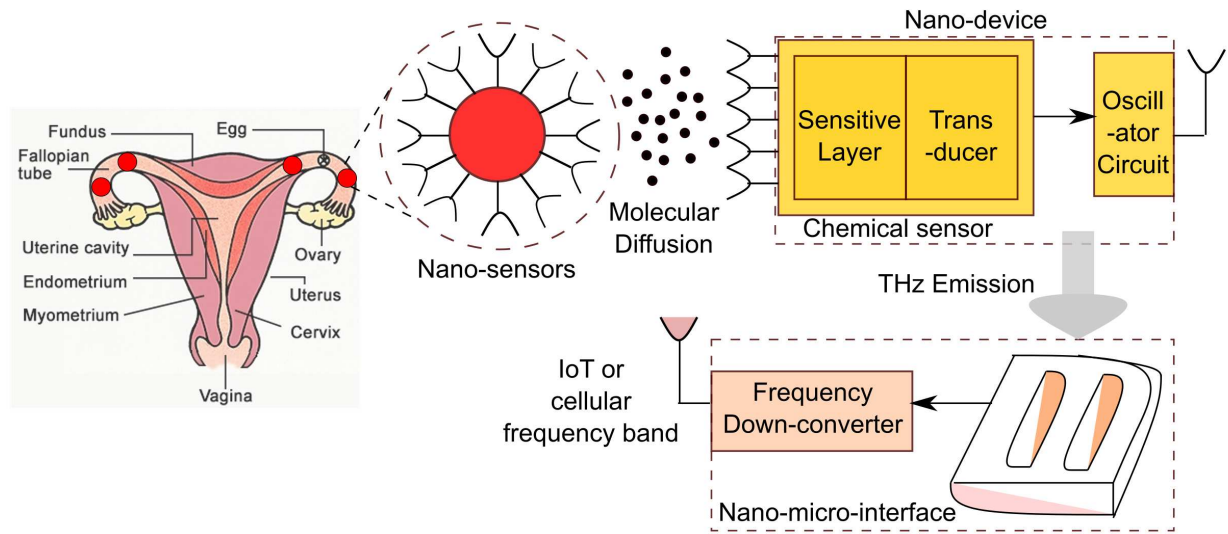}
\vspace{-165mm}
\caption{Combining molecular and electromagnetic wave-based communication for monitoring fertility within female reproductive system.}
\label{fig:Fallopian1}
\vspace{-4mm}
\end{figure*}

The nano-device acts a transceiver composed of a combined biological receiver and an electromagnetic (EM) signal transmitter. Specifically, it consists of a chemical nano-sensor that can detect transmit molecule concentration from the nano-sensors and can convert this concentration information to an electrical signal. The electrical signal is used to trigger an electromagnetic signal in the Terahertz (THz) frequency band containing the information on the presence of eggs in the Fallopian tubes. This signal is, in turn, transmitted to a dual-antenna (THz and over-the-air backhaul IoT or cellular network frequency range) nano-micro interface that is deployed on the human body surface. An info-graphic representation of communication involving the intra-body nano-sensors and nano-device and on-body nano-micro-interface is presented in Fig.~\ref{fig:Fallopian1}.

Nano-micro-interface forwards relevant information on the presence/absence of eggs in the Fallopian tube to the individual's handheld device, typically a smart phone. The handheld device forwards this information to the gateway for the consultant physician, audio-video communication and generating alerts for specific actions (like inducing fertilization, artificial insemination etc.). At the same time, the information received from a monitored individual is continuously forwarded to a network server for maintaining database in case of further use. 

Our proposed overall architecture can therefore be grouped into three communication modules operating over different media and modes.
\begin{itemize}
\item{Communication Module-1: The nano-network comprising of the intra-body nano-sensors/nano-nodes and nano-transmitters/actuators deployed within the Fallopian Tube, operating using assisted diffusion-based molecular communication.}
\item{Communication Module-2: The nano-network between the intra-body nano-transmitters/actuators and wearable or on-body device, communicating using THz band}
\item{Communication Module-3: The conventional IoT backhaul network for the data transmission between the wearable/on-body device to the remote sink and/or handheld device using 4G/5G or WiFi.}
\end{itemize}
It is to be noted that the proposed architecture is based on different links with significantly diverse characteristics; the transmission media for in-body environment will be blood, tissue, and other body fluids, whereas the on-body device shall be transmitting information over the air.

\section{Communication Module-1}
Communication Module-1 consists of the nano-nodes deployed within the Fallopian tube and they communicate among themselves through assisted diffusion based molecular communication (MC). The major underlying assumption for such communication is the possibility of successful transmission of signal/molecule from the transmitter to receiver, while catering to the losses. There needs to be favorable alignment of the living cells so the molecules from transmitting unit may travel. It is also assumed that the sensors which could detect the application-specific bio-marker is available and it could timely detect/report to the transmitter unit. Besides to preserve the health of human tissue, the field strength of signals, average time of exposure and power density must remain within the exposure limits set by standards such as IEEE C95.1. The restrictions have to be applied also to the continuous wave or short pulse-based modulation techniques used for transmission of signals between the nano-nodes.

Various technologies for establishing MC within the intra-body environment have been proposed considering the size and scale of the molecules. Calcium ions have often been proposed as information carriers, with approximate size ranging between 100 and 200 pm and capacity to move up to distances of 300 {\textmu}m; these ions can only achieve velocity of 30 {\textmu}m/sec \cite{kulak}. Moreover, polymers and molecular motors (wit size 100 nm) have also been used for MC. Since the movement of ions is very slow naturally, active assisted diffusion has been proposed to transmit the information by encoding bits in DNA strands and using platinum and gold nano rods for participating in the chemical reactions. Neurons/Nerve cells are used as a channel to facilitate the movement of molecules; these cells may range from a size of few micrometers to 1 or 2 m, i.e, considering the nerve that goes from toe to brain. The impulses (electric/action potential) resulting from the movement of neurotransmitters travel very fast through the neuron and may acquire speed up to 120 m/s. However, the overall throughput for the communication within neurons is limited because of the idle time after the propagation of action potential once, which is at least up to 1 ms. 

Unlike traditional RF communication systems, information is modulated on the type/structure, duration of emission or number of released molecules in case of MC. Common techniques used for assisted diffusion based MC are On-Off (1, if a finite number is the molecular concentration and 0, otherwise over the bit duration), Multilevel amplitude modulation (molecular concentration is encoded on amplitude and frequency of a continuous sinusoidal/cosine wave), Concentration shift keying (number of information molecules released is considered as the signal amplitude) and Molecular shift keying (the different types/structures of the information molecules is considered as the signal amplitude) \cite{yang}.

On the receiver side, few of the transmitted molecules activate some receptors on the surface of the receiver and the number of activated receptors constitute the receive signal. Different models have been proposed for describing the reception phenomenon based on the kind and number of received molecules. The receivers can be passive (concentration of the received molecules within the receiver determines the received signal), ligand binding (concentration of output molecules produced by reversible reaction between transmit molecules or ligands and the receptors determine the received signal), fully-absorbing (concentration of the transmitted molecules fully absorbed by a sphere-like receiver determines the received signal) and reversible absorption (first-order reversible reaction-diffusion equation is used to model the interaction of the molecules absorbed within the receiver sphere) \cite{akyildiz}.

Signals travelling over the molecular diffusion-based communication channel are affected by noise inherently due to the random arrival of the molecules at the receiver and owing to external sources like thermal, physical, sampling and counting noise. Similar to traditional wireless communication system, many coding techniques have been proposed to improve reliability in such a scenario. Such techniques include Hamming, minimum-energy, Euclidean geometry-based low density parity check and cyclic Reed-Muller codes \cite{lu}. 

\section{Communication Module-2}

Communication Module-2 comprises of the nano-nodes receiving information from the Fallopian tubes and forwarding that information to a nano-micro on-body interface. Transfer of information over this section is enabled through Terahertz (THz) electromagnetic (EM) wave based communications \cite{alomainy2019nano}. Since scattering pathloss is negligible relative to absorption loss, THz is preferred over short-distance intra-body to body-surface communication. Graphene-based antennas that operate at the THz (0.1 to 10 THz) band are used on both the in-body and body-surface devices \cite{yang}.  

Although THz band offers high data rate of up to Tbps for the intra-body communication, but there are several challenges due to the unique lossy nature of human body tissue as a communication channel. Not only the signals face high frequency attenuation, but the communication range is also limited to few millimeters only; even during this short-range transmission, path loss as high as 120 dB is expected due to spreading wavefront, frequency-selective absorption and conversion of EM energy into kinetic. 

In order to cater to the limitations in size and power of the nano-devices carrier-less pulse based modulation technique like time-spread on-off keying (TS-OOK) has been recommended. In TS-OOK, femto-second long pulses spread in time are used for exchange. The pulse represents a logical `0' if the transmitter is silent and a logical `1' otherwise. Similarly, energy efficient fixed-length code-words with constant weights has been proposed for enhancing reliability for this section of communication \cite{jornet}.

Several medium access control (MAC) protocols have also been proposed for energy efficient and interference-free data transmission over the THz in-body link. Initially, the MAC schemes were designed based on the assumption of single transmitter and receiver; however, the emerging applications with increasing number of nodes would require more sophisticated MAC designs. IB-MAC has been developed for facilitating high data rate EM communications within intra-body environment \cite{ghafoor}. It is a time division multiple access (TDMA) based protocol which utilizes superframe structure and can support up to 100 intra-body nodes. Although the concept of node priority has been introduced, latency could pose serious risks due to using conventional TDMA scheme. PHLAME was also developed for intra-body nano networks and was based on the principle of mutual decision making between transmitter and reciever about the transmission parameters.


\section{Communication Module-3}

Communication Module-3 consists of router nodes, central gateway and end-nodes, that route information from the on-body interface to a central platform (gateway), manages the received information and then forwards it to the relevant application endpoint and takes appropriate action respectively. Communication takes place over the air using RF waves and existing Internet-of-Things (IoT) and cellular backhaul network. The available network can be infrastructure-based or ad-hoc-based. In infrastructure-based set-up, information from on-body device is sent directly to the gateway and in ad-hoc-based set-up, information is routed through multiple access points (APs) to the gateway. The APs are chosen in an ad-hoc basis depending on the location of the monitored individual and the corresponding environmental conditions, like instantaneous signal-to-noise ratio (SNR). 

Some perspective design technologies can be proposed for energy-efficient, low-latency and high throughput transfer of information for this section of network. Large scale Multiple-input-multiple-output (MIMO) employ hundreds of antennas for serving typically a few dozen terminals while sharing the same time-frequency resources. Therefore, it is possible to employ multiple antennas at the IoT gateway, APs and the end-nodes to enable energy-efficient transmission over the wireless environment suffering from deep fading and shadowing \cite{rossi}. It is worth-mentioning here that this section of the communication network is designed depending on the application at hand. The application end-point can also serve as a database where continuous monitoring data is stored.

Existence of multiple users will lead to increased utilization of scarce spectrum. Dynamic spectrum management as well as energy efficient and environment-aware design can be jointly considered in such a scenario. A framework combining distributed decision fusion at the IoT gateway and orthogonal frequency division multiplexing (OFDM) based collaborative wideband spectrum sensing, similar to the one proposed in \cite{dey2} can be used. Just like spectrum, existence of multiple users will lead to increased usage of other resources like energy, computing memory, data processing tools etc. In order to handle and integrate computing resources in networks like the one proposed here, distributed artificial intelligent (AI)-based methods can be investigated and implemented.

\section{Simulation Results}

In this section, we present preliminary simulation results to evaluate the performance of proposed architecture. As time is the most crucial parameter for our envisaged application, we simulate information rates achievable over the distance between the actual monitored individual and the terminal of the consultant physician. 

For communication module-1, we use the particle-based stochastic simulator of \cite{schober} with 5 nano-sensors deployed uniformly on the circumference of a circle of radius 45~nm and the nano-device (biological transceiver) put at the center of the circle. All the sensors release the same number of molecules (i.e. $10^4$) with diffusion coefficient of $4.365 \times 10^{-10}$~$m^2$/s over an environment with constant temperature and viscosity. If eggs are detected in the Fallopian tube, a binary `1' is transmitted by emitting impulses of molecules. A binary sequence of 50 bits is transmitted by each sensor and simulated information rate is averaged over $10^5$ random realizations of the sequence.  At the nano-receiver, samples of the molecular impulse are collected over a bit duration of 400~$\mu$s and 10 samples are collected each 50 sec. 

For the communication module- 2, electromagnetic signal generated by the nano-transceiver is transmitted to the Nano-micro-interface using TS-OOK modulated 100 femtosecond long pulses. It is worth-noting here that the time between symbols is much longer than the symbol duration and the ratio of the time interval between symbols to symbol duration is assumed to be 100. The frequency of operation is 0.5 THz to 1.5 THz with a bandwidth of 1 THz.

For communication module-3, we include a synthetic IoT framework with one handheld device (with 1 antenna), gateway (16 antennas) equipped with the cloud computing platform and relevant endpoint (each with 8 antennas), like network server and physician's terminal. The channel between the individual, gateway and endpoint is assumed to be dispersive with Rayleigh fading envelope and normalized Doppler frequency of 0.01 Hz while the pathloss is experienced according to the 3GPP Urban Micro-cell (UMi) model \cite{umi}. The application endpoints use multiple-input-multiple-output (MIMO) - zero forcing (ZF) based symbol decoder for detecting information received from the monitored individual. 

\begin{figure}[t]
\begin{center} 
 \includegraphics[height=4cm, width=8cm]{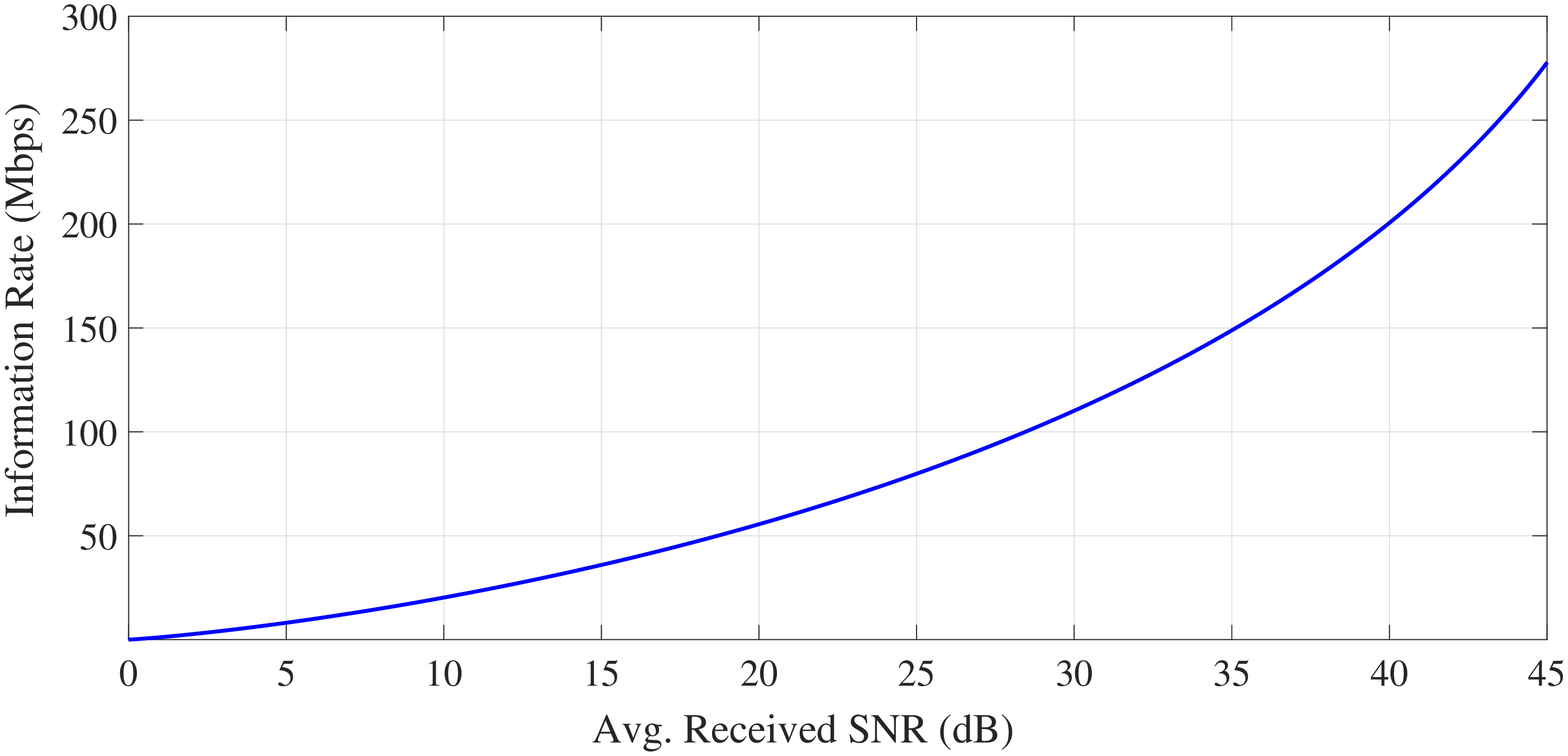}
\end{center}
\caption{Information rate as a function of the average SNR over the end-to-end link from the monitored individual to the application endpoint.}
  \label{info_snr}
\end{figure}

\begin{figure}[t]
\begin{center} 
 \includegraphics[height=6cm, width=8cm]{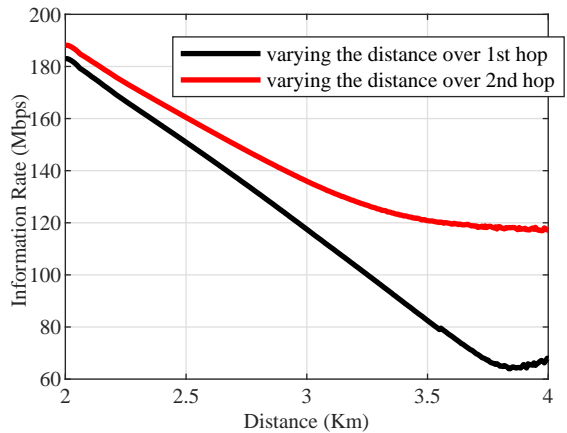}
\end{center}
\caption{Information rate as a function of the distance over each hop; distances over the first and second hops are kept fixed at 1 Km for the red and black curves respectively.}
  \label{info_dist}
\end{figure}

Fig.~\ref{info_snr} studies the information rate achievable as a function of the average signal-to-noise ratio (SNR) experienced over the end-to-end link. The distances between the individual and gateway and the gateway and application endpoint are both kept constant at 1 Km. A high data rate of 300 Mbps is achievable with this multi-mode (molecular communication and electromagnetic waves) transfer of information. Fig.~\ref{info_dist} studies the change in information rate with the increase in distance over each hop. The first hop involves the link between the individual and the gateway, whereas the second hop refers to the link between the gateway and the endpoint. The average SNR over the end-to-end link is kept fixed at 15 dB. It is quite evident that increase in distance between the individual and the gateway impacts the overall information rate more than the increase over the second hop.

\section{Challenges and Future Direction}
The domain of intra-body communication is undoubtedly a futuristic technology which promises to reduce the healthcare cost and significantly enhance the quality of healthcare services, mainly due to facilitating early diagnosis.  The communication architecture proposed in this work will enable the medical researchers and gynaecologists to monitor the fertility of their patients. It is expected that when the patients will be enabled to detect the presence of eggs in the Fallopian tubes, their chances for natural conception will increase. This would clearly result in reduced healthcare burden and suffering associated with the artificial insemination (IUI) and In-vitro fertilization (IVF) methods. In addition to helping to identify the presence of eggs, once fertilized, the nano-sensors could also be used to track the location and state of egg as they transport and implant to the endometrium. Furthermore, the proposed architecture shall not only aid the females during their journey of conception, but will also facilitate the monitoring of Fallopian activity in general for diagnosis of cysts, irregular cycles, and even cancers at early stages. However, highly application specific nature of intra-body communication makes it challenging to design and develop the solutions for hundreds of crucial human health issues. Some of these challenges are listed below:

\begin{itemize}
    \item Although the THz communication link will increase the chances of early detection of egg presence, the designers will have to take care of the intricate considerations such as node size, antenna directionality, path loss and scattering of molecules, line of sight blockage and mobility of signals. 
    \item The use of Graphene based antennas and THz communication has already been proved to not posing health risks for the patients, it will still be crucial to conduct sophisticated clinical trials in order to avoid any possible future health risk for the patient and her fertility status.
    \item Although the interference and packet collisions are expected to be lower in the Fallopian tubes as compared to the conventional networks, there will be a need to develop error control MAC mechanisms in order to reduce the probability of re-transmissions which could otherwise be high due to the path and absorption losses.
    \item In addition to the physical design of communication devices, there is a need to develop communication protocols for entire stack based on compatibility with the unique transmission medium of Fallopian tube environment.
    \item The ethical constraints associated with conducting clinical trials could also act as a major barrier against the proposed solution. Since this is a novel proposal of implanting nano-sensors inside the Fallopian tubes, ethical concerns and clinical approvals will be required to begin the clinical trials. 
\end{itemize}

\section{Conclusions}
Based on the recent evolutions in nano-sensors and intra-body communication technologies, this letter has proposed a novel communication architecture to timely detect the presence of eggs in the female Fallopian tube. The implanted sensor nodes within the Fallopian tube shall communicate with the on-body device, hand-held device and  remote locations to transmit alerts. It is expected that the proposed architecture shall enable the users and gynaecologists to effectively monitor the fertility state which shall in turn improve the rate of natural conception.


\end{document}